# Bayesian analysis of dynamic binary data: A simulation study and application to economic index SP

Ali Reza Fotouhi
University of the Fraser Valley
33844 King Road, Abbotsford
BC, V2S 7M8,Canada
E-mail: ali.fotouhi@ufv.ca
Fax: +1 (604) 855-7614

**Abstract:**

It is proposed in the literature that in some complicated problems maximum likelihood estimates (MLE) are not suitable or even do not exist. An alternative to MLE for estimation of the parameters is the Bayesian method. The Markov chain Monte Carlo (MCMC) simulation procedure is designed to fit Bayesian models. Bayesian method like classical method (MLE) has advantages and disadvantages. One of the advantages of Bayesian method over MLE method is the ability of saving the information included in past data through the posterior distributions of the model parameters to be used for modelling future data. In this article we investigate the performance of Bayesian method in modelling dynamic binary data when the data are growing over time and individuals.

## 1. Introduction

Maximum Likelihood Estimation (MLE) is a method of estimating the parameters of a statistical model. In general, for a fixed set of data and underlying statistical model, the method of maximum likelihood selects values of the model parameters that produce a distribution that gives the observed data the maximum probability. MLE gives a unified approach to estimation in the case of the normal distribution and many other problems. In some complicated problems, MLEs are not suitable or do not exist. Even for some simple and popular cases, like logistic regression models, MLE needs some conditions for existence (Albert and Anderson, 1984). The maximum likelihood estimator has essentially no optimal properties for finite samples and is sensitive to the initial values. However, the maximum likelihood estimator possesses a number of attractive asymptotic properties, for many problems. These asymptotic properties include consistency, asymptotic normality, and efficiency.

Bayesian method is an alternative to the classical method (MLE) for estimation of the parameters in a statistical model. Bayesian method treats parameters as unknown random variables, and it makes inferences based on the posterior distributions of the parameters. The Markov chain Monte Carlo (MCMC) simulation procedure is designed to fit Bayesian models. It follows from Bayes' theorem that a posterior distribution is the product of the likelihood function and the prior distribution of the parameter. Except in the simplest cases, it is very difficult to obtain the posterior distribution directly and analytically. Bayesian modelling relies on simulations to generate sample from the desired posterior distribution and use the simulated draws to approximate the distribution and to make all of the inferences.

Bayesian method using MCMC procedure like classical method (MLE) has advantages and disadvantages, and there are some similarities. When the sample size is large, MCMC often provides results for parametric models that are very similar to the results produced by MLE. One advantage of using Bayesian analysis is that it provides a natural and principled way of combining prior information with data (Berger 1985). We can incorporate past information about a parameter and form a prior distribution for future analysis. When new observations become available, the previous posterior distribution can be used as a prior. It provides inferences that are conditional on the data and are exact, without reliance on asymptotic approximation. It is assumed that small sample inference proceeds in the same manner as if one had a large sample and obeys the likelihood principle. Classical inference (MLE) does not in general obey the likelihood principle and provides a convenient setting for a wide range of models, such as hierarchical models and missing data problems. MCMC method along with other numerical methods, makes computations tractable for all parametric models.

There are also disadvantages in using Bayesian analysis. Firstly, it does not tell us how to select a prior. If we do not proceed with caution, we may generate misleading results. Secondly, it often comes with a high computational cost, especially in models with a large number of parameters. Simulations in MCMC provide slightly different answers unless the same random seed is used. Slight variations in simulation results do not contradict the claim that Bayesian inferences are exact. The posterior distribution of a parameter is exact, given the likelihood function and the priors, while simulation-based estimates of posterior quantities can vary due to the random number generator used in the procedures. See Berger (1985) for more discussion on advantages and disadvantages of MCMC.

PROC MCMC in SAS is a procedure that is suitable for fitting a wide range of Bayesian models. To use the procedure, we need to specify a likelihood function for the data and prior distributions for the parameters. We might also need to specify hyper-prior distributions if we are fitting hierarchical models. PROC MCMC obtains samples from the corresponding posterior distributions and produces summary and diagnostic statistics. We can analyze data that have any programmable likelihood, prior, or hyper-prior with PROC MCMC. The default algorithm that PROC MCMC uses is an adaptive blocked random walk Metropolis algorithm that uses a normal proposal distribution.

In many applications data are massive and are continuously collected over time. There is always a possibility of missing part or even the entire of data. One strategy is to use as much data as are available using MLE or Bayesian method. In this case we lose the information included in missed data. An advantage of Bayesian method using MCMC procedure is the possibility of saving the information included in the past data into the posterior distribution of the parameters and use them in the analysis of current data. We discuss this property in a logistic regression model through simulation study and application to economic index SP. We consider time-constant and time varying explanatory variables.

**2. Model and estimation method**

In this section we introduce a logistic model for analyzing binary data. Consider a binary data set observed over time consisting of a response $y_{ij}$ for the $j^{th}$ observation of the $i^{th}$ independent individual and a $p \times 1$ vector $x_{ij}$ of covariates associated with the response $y_{ij}$. Let $\beta$ be a $p \times 1$ vector of unknown fixed effect parameters associated with $x_{ij}$. A logistic model of the following form can describe the binary data of this type.

$$p(y_{ij} = 1 | x_{ij}) = \frac{\exp(x'_{ij}\beta)}{1+\exp(x'_{ij}\beta)} \quad \text{and} \quad p(y_{ij} = 0 | x_{ij}) = \frac{1}{1+\exp(x'_{ij}\beta)}$$
$$i = 1, 2, ..., I \; ; \; j = 1, 2, ..., n_i \tag{1}$$

For binary data the relation between mean and variance, $\text{var}(y) = E(y)(1 - E(y))$, is often inconsistent with empirical evidence. Therefore the logistic model with the linear predictor $x'_{ij}\beta$, which does not account for over-dispersion or under-dispersion, would not be appropriate for analyzing the longitudinal binary outcomes. Omitted variables from the linear predictor may be the main reason for the over-dispersion or under-dispersion and may have different effects on the linear predictor for different individuals. To accommodate over-dispersion or under-dispersion, we add a random variable to the linear predictor, which leads to the well-known random effects model. The conditional probability function for individual $i$ using a logistic model with $\varepsilon_i$ as the individual specific error term with mean zero and standard deviation $\sigma$ in the linear predictor is of the following form

$$p(y_{ij} = 1 | x_{ij}) = \frac{\exp(x'_{ij}\beta + \varepsilon_i)}{1+\exp(x'_{ij}\beta + \varepsilon_i)} \quad \text{and} \quad p(y_{ij} = 0 | x_{ij}) = \frac{1}{1+\exp(x'_{ij}\beta + \varepsilon_i)}$$
$$i = 1, 2, ..., I \; ; \; j = 1, 2, ..., n_i \tag{2}$$

The MCMC method is a general simulation method for sampling from posterior distributions and computing posterior quantities of interest. MCMC method take sample successively from a target distribution and each sample depends on the previous one (the Markov chain property). A Markov chain then is a sequence of random variables, $\beta_0, \beta_1, \beta_2, ...$ for which the random variable $\beta^t$ depends on $\beta^0, \beta^1, \beta^2, ..., \beta^{t-1}$ only through $\beta^{t-1}$. Monte Carlo method, as in Monte Carlo integration, is used to approximate an expectation of the produced Markov chain samples. The earliest reference to MCMC simulation occurs in the physics literature. Metropolis and Ulam (1949) and Metropolis et al. (1953) describe what is known as the Metropolis algorithm. The algorithm can be used to generate sequences of samples from the joint distribution of multiple variables, and it is the foundation of MCMC. Hastings (1970) generalized their work, resulting in the Metropolis Hastings algorithm. Geman and Geman (1984) analyzed image data by using what is now called Gibbs sampling. The Metropolis algorithm is simple and practical, and it can be used to obtain random samples from any arbitrarily complicated target distribution of any dimension. The Bayesian procedures use a special case of the Metropolis algorithm called the Gibbs sampler to obtain posterior samplers. Suppose we want to obtain $T$ samples from a distribution with probability density function $g(\beta | y)$. Suppose $\beta^t$ is the $t^{th}$ sample from $g(\beta | y)$. To use the Metropolis algorithm, we need to have an initial value $\beta^0$ and a symmetric density $q(\beta^{t+1} | \beta^t)$. For the $(t+1)^{th}$ iteration, the algorithm generates a sample from $q(. | .)$ based on the current sample $\beta^t$, and it makes a decision to either accept or reject the new sample. If the new sample is accepted, the algorithm repeats itself by starting at the new sample. If the new sample is rejected, the algorithm starts at the current point and repeats. The algorithm is self-repeating, so it can be carried out as long as required. In practice, we have to decide the total number of samples needed in advance and stop the sampler after that required iterations have been completed. Procedure MCMC in SAS uses this method for the estimation of the parameters.

### 3. Simulation study

In this section we set up a simulation study to generate longitudinal binary data for different sample sizes and different level of heterogeneity. We investigate the effects of sample size, level of heterogeneity, and the effect of prior distributions on the parameters estimation. We consider $I(=100,200,500,1000)$ individuals over $T = (12,100)$ periods of time. We assume a low to high heterogeneous longitudinal binary data with standard deviation ($\sigma = 0.5, 1, 2$). We consider both time-constant and time varying explanatory variables $x_1$ and $x_2$ with true effects of $\beta_1 = 1$ and $\beta_2 = 1$ respectively. We assume the constant term to be $\beta_0 = -1$. We generate $N = 30$ samples each includes $I$ individuals over $T$ periods of time according to the following steps.

**Step 1:** Generate random effects $\varepsilon_i$ from a normal distribution with mean zero and variance $\sigma^2$.

**Step 2:** For time-constant explanatory variable generate $x_{1ij}$ from Bernoulli probability distribution with probability of success $0.5$. For time-varying explanatory variable, generate $x_{2ij}$ from Nerlove (1971) process

$x_{2ij} = .1j + .5x_{2i,j-1} + uniform(-.5,.5)$ with $x_{2i1} = uniform(-.5,.5)$. This time series is well recognized for age trend.

**Step 3:** Calculate $\mu_{ij} = \beta_0 + \beta_1 x_{1ij} + \beta_2 x_{2ij} + \varepsilon_i$ and generate $y_{ij}$ from Bernoulli probability distribution with $p_{ij} = \frac{\exp(\mu_{ij})}{1+\exp(\mu_{ij})}$.

**Step 4:** repeat steps 2 and 3, in order, for $j = 1,2,...,T$.
**Step 5:** Repeat steps 1 to 4 for $i = 1,2,...,I(=100,200,500,1000)$.
**Step 6:** repeat steps 1 to 5 for $N = 30$ times.

These steps generate 30 matrices of data $M$ each having $I$ rows and $T$ columns. We partition each one of these matrices into four matrices each having $\frac{I}{2}$ rows and $\frac{T}{2}$ columns as

$$M = \begin{bmatrix} M_{11} & M_{12} \\ M_{21} & M_{22} \end{bmatrix}$$

We use Procedure MCMC from SAS 9.4 to fit the model introduced in sections 2 to different combinations of the generated data sets. For all runs with uninformative prior distributions, we assume *normal (Mean=0, Variance=10000)* for the parameters $\beta_0, \beta_1, \beta_2$ and *igamma (Shape=0.001, Scale=0.001)* for $\sigma^2$. For runs with informative prior distributions we used the posterior distributions of the Mean, Variance, Shape and Scale parameters obtained from fitting the first part of data. In order to investigate the effectiveness of using informative prior distributions we consider six runs based on the following conditions.

**R1:** We fit the data included in $[M_{11} \quad M_{12}]$ assuming uninformative prior distributions for the parameters and use the obtained posterior distributions of the parameters to fit the data included in $[M_{21} \quad M_{22}]$.

**R2:** We fit the data included in $\begin{bmatrix} M_{11} \\ M_{21} \end{bmatrix}$ assuming uninformative prior distributions for the parameters and use the obtained posterior distributions of the parameters to fit the data included in $\begin{bmatrix} M_{12} \\ M_{22} \end{bmatrix}$.

**R3:** We fit the data in $M_{11}$ assuming uninformative prior distributions for the parameters and use the obtained posterior distributions of the parameters to fit the data included in $M_{22}$.

**R4:** We fit the data in $M_{22}$ assuming uninformative prior distributions for the parameters for comparison.

**R5:** We fit the data in $[M_{21} \quad M_{22}]$ assuming uninformative prior distributions for the parameters for comparison.

**R6:** We fit the data in $\begin{bmatrix} M_{12} \\ M_{22} \end{bmatrix}$ assuming uninformative prior distributions for the parameters for comparison.

In order to compare the estimates obtained from comparable runs we use Mean Square Error (MSE) which includes both bias and variance of the estimate.

$$MSE(\hat{\beta}) = (\hat{\beta} - \beta)^2 + var(\hat{\beta})$$

The results from this simulation study are reported in Tables 1 to 12.

Comparing runs R1 and R5 shows that using informative priors obtained from first half of data, collected over individuals, produces smaller MSE for $\beta_1$ and $\beta_2$ when sample size is 500 or 1000 and $\sigma = 0.5$. But using informative priors produces smaller MSE for $\beta_1$ and $\beta_2$ for any sample size when there is large level of heterogeneity ($\sigma = 1$ or 2).

Comparing runs R2 and R6 shows that using informative priors obtained from first half of data collected over time produces smaller MSE for $\beta_1$ and $\beta_2$ for any sample size and any value of $\sigma$. This is mostly due to the smaller standard deviation of the estimate produced by informative priors.

Comparing runs R3 and R4 indicates that uninformative priors produce smaller MSE for $\beta_1$ and $\beta_2$ for any sample size and any value of $\sigma$. This could be due to the fact that the data ($M_{22}$) used in run R3 are collected over new individuals and times. As in some applications the data window is large we have repeated runs R2 and R6 assuming *T=100, I=100, N=30,* and $\sigma = 1$. The results are shown in Table 13. The results reported in Table 13 indicate that the uninformative prior distributions of the parameters produce very biased estimate for the parameters with large MSEs. The informative prior distributions obtained from the first half of data produce unbiased estimates for $\beta_0$, $\beta_1$, and $\sigma$. The effect of time-varying explanatory variable, $\beta_2$, is estimated with a little bias but all MSEs are very small as compare to the MSEs obtained from using uninformative priors. We have checked the performance of MCMC procedure and have found no warning or error in any run. It seems that the informative priors are more effective than uninformative priors when time-varying explanatory variable is in the model and data window is large.

## 4. Application

In order to investigate if the results from simulation study are consistent with real life applications we applied the proposed logistic model to economical index SP recorded from 1960 to 2018. We have considered the linear predictor $\beta_0 + \beta_1(time - 1960)$ where time is colander time in year. The response is the binary variable $y = 1$ if the return value exceeds 1.4 and $Y = 0$ if the return value does not exceeds 1.4. The threshold 1.4 is borrowed from the articles by Fotouhi (2019) and Gilli. M, KÄellezi, (2006). They used this threshold for analyzing the extreme values of SP index using Peak-Over Threshold method. We applied the logistic regression model to data from 1960 to 2004 using uninformative prior distributions and used the obtained posterior distributions of the parameters to fit the data from 2005 to 2018. The result are shown in Table 14. The time effect, $\beta_1$, is estimated significantly negative when uninformative priors are used while it is estimated significantly positive when informative priors are used. The positive estimate of $\beta_1$ is consistent with the increasing empirical values of the odds of success ($y = 1$) over time while a negative estimate is not. The standard deviation of the random effects is estimated significantly positive which indicates that the logistic model could capture the heterogeneity of the data. The result from this application is consistent with the simulation result and indicates the usefulness of using informative prior distributions for dynamic binary data obtained from the economical index SP.

**5. Conclusion**

We have performed a simulation study to show the importance of using Bayesian approach in fitting dynamic binary data in which the data are growing over time and individuals. The objective of this research is showing that saving the information included in current data through the posterior distributions of the parameters and use them to fit the future data produces better estimates for the structural parameters of the logistic regression. This approach could be useful in many applications that data are massive and are continuously collected over time and there is a possibility of missing part or even the entire of past data.

We have considered time-constant and time-varying explanatory variables and simulated data from low to high level of heterogeneity for different sample sizes. Our simulations show that, when data are growing over time for a fixed number of individuals, using informative prior distributions for the structural parameters of the model, obtained from past data, fit the future data with smaller mean square error than using uninformative prior distributions for any sample size and any level of heterogeneity. When data are growing over individuals for fixed period of time, we reached to the same conclusion except for the case that the level of heterogeneity is low and the sample size is small. Our simulations show that informative prior distributions of the parameters obtained from past data are not better than uninformative prior distributions when current data are collected over new individuals and times.

In order to investigate if the results from simulation study are consistent with real life applications we applied the proposed logistic model to economical index SP recorded from 1960 to 2018. We found the result of this application consistent with the simulation result and indicates the usefulness of using informative prior distributions for dynamic binary data obtained from the economical index SP

**Table 1:** Estimate of $\beta_0$ with true value of -1. Standard deviation of the random effects is 0.5.

| Run | N | Mean | SD | LCL | UCL | MSE |
|---|---|---|---|---|---|---|
| R1 | 100 | -1.068 | 0.157 | -1.127 | -1.009 | 0.029 |
|    | 200 | -0.978 | 0.138 | -1.030 | -0.927 | 0.020 |
|    | 500 | -0.995 | 0.078 | -1.024 | -0.966 | 0.006 |
|    | 1000 | -0.996 | 0.058 | -1.018 | -0.975 | 0.003 |
| R2 | 100 | -1.023 | 0.207 | -1.101 | -0.946 | 0.043 |
|    | 200 | -0.958 | 0.139 | -1.010 | -0.906 | 0.021 |
|    | 500 | -0.959 | 0.084 | -0.990 | -0.927 | 0.009 |
|    | 1000 | -0.937 | 0.055 | -0.958 | -0.917 | 0.007 |
| R3 | 100 | -1.049 | 0.246 | -1.141 | -0.957 | 0.063 |
|    | 200 | -0.936 | 0.210 | -1.014 | -0.857 | 0.048 |
|    | 500 | -0.970 | 0.107 | -1.010 | -0.930 | 0.012 |
|    | 1000 | -0.959 | 0.078 | -0.988 | -0.930 | 0.008 |
| R4 | 100 | -1.026 | 0.132 | -1.075 | -0.976 | 0.018 |
|    | 200 | -0.981 | 0.118 | -1.025 | -0.937 | 0.014 |
|    | 500 | -0.987 | 0.063 | -1.011 | -0.963 | 0.004 |
|    | 1000 | -0.978 | 0.042 | -0.994 | -0.963 | 0.002 |
| R5 | 100 | -1.018 | 0.207 | -1.095 | -0.941 | 0.043 |
|    | 200 | -1.019 | 0.138 | -1.071 | -0.968 | 0.019 |
|    | 500 | -0.995 | 0.099 | -1.032 | -0.958 | 0.010 |
|    | 1000 | -0.970 | 0.062 | -0.993 | -0.947 | 0.005 |
| R6 | 100 | -0.994 | 0.357 | -1.127 | -0.860 | 0.128 |
|    | 200 | -0.985 | 0.286 | -1.092 | -0.878 | 0.082 |
|    | 500 | -0.957 | 0.164 | -1.019 | -0.896 | 0.029 |
|    | 1000 | -0.963 | 0.115 | -1.006 | -0.920 | 0.015 |

**Table 2:** Estimate of $\beta_0$ with true value of -1. Standard deviation of the random effects is 1.

| Run | N | Mean | SD | LCL | UCL | MSE |
|---|---|---|---|---|---|---|
| R1 | 100 | -1.040 | 0.173 | -1.104 | -1.104 | 0.031 |
|    | 200 | -0.971 | 0.114 | -1.014 | -1.014 | 0.014 |
|    | 500 | -0.980 | 0.085 | -1.012 | -1.012 | 0.008 |
|    | 1000 | -0.984 | 0.063 | -1.008 | -1.008 | 0.004 |
| R2 | 100 | -1.001 | 0.225 | -1.085 | -1.085 | 0.051 |
|    | 200 | -0.951 | 0.132 | -1.000 | -1.000 | 0.020 |
|    | 500 | -0.946 | 0.091 | -0.980 | -0.980 | 0.011 |
|    | 1000 | -0.935 | 0.069 | -0.961 | -0.961 | 0.009 |
| R3 | 100 | -1.005 | 0.289 | -1.112 | -0.897 | 0.084 |
|    | 200 | -0.918 | 0.198 | -0.992 | -0.844 | 0.046 |
|    | 500 | -0.969 | 0.126 | -1.016 | -0.923 | 0.017 |
|    | 1000 | -0.954 | 0.092 | -0.988 | -0.920 | 0.011 |
| R4 | 100 | -1.021 | 0.165 | -1.083 | -1.083 | 0.028 |
|    | 200 | -0.978 | 0.107 | -1.018 | -1.018 | 0.012 |
|    | 500 | -0.986 | 0.073 | -1.013 | -1.013 | 0.005 |
|    | 1000 | -0.977 | 0.053 | -0.997 | -0.997 | 0.003 |
| R5 | 100 | -0.999 | 0.295 | -1.109 | -1.109 | 0.087 |
|    | 200 | -1.029 | 0.175 | -1.095 | -1.095 | 0.031 |
|    | 500 | -0.991 | 0.115 | -1.034 | -1.034 | 0.013 |
|    | 1000 | -0.973 | 0.084 | -1.004 | -1.004 | 0.008 |
| R6 | 100 | -0.984 | 0.372 | -1.227 | -1.227 | 0.139 |
|    | 200 | -0.972 | 0.272 | -1.074 | -1.074 | 0.075 |
|    | 500 | -0.954 | 0.166 | -1.016 | -1.016 | 0.030 |
|    | 1000 | -0.924 | 0.131 | -0.973 | -0.973 | 0.023 |

**Table 3:** Estimate of $\beta_0$ with true value of -1. Standard deviation of the random effects is 2.

| Run | N | Mean | SD | LCL | UCL | MSE |
|---|---|---|---|---|---|---|
| R1 | 100 | -1.016 | 0.256 | -1.111 | -0.920 | 0.066 |
|    | 200 | -0.956 | 0.163 | -1.017 | -0.895 | 0.029 |
|    | 500 | -0.985 | 0.140 | -1.037 | -0.933 | 0.020 |
|    | 1000 | -0.978 | 0.098 | -1.015 | -0.941 | 0.010 |
| R2 | 100 | -1.011 | 0.351 | -1.142 | -0.880 | 0.123 |
|    | 200 | -0.960 | 0.180 | -1.027 | -0.893 | 0.034 |
|    | 500 | -0.961 | 0.147 | -1.016 | -0.906 | 0.023 |
|    | 1000 | -0.916 | 0.112 | -0.957 | -0.874 | 0.020 |
| R3 | 100 | -1.068 | 0.416 | -1.223 | -0.912 | 0.178 |
|    | 200 | -0.912 | 0.281 | -1.017 | -0.807 | 0.087 |
|    | 500 | -0.973 | 0.178 | -1.040 | -0.907 | 0.032 |
|    | 1000 | -0.970 | 0.146 | -1.024 | -0.916 | 0.022 |
| R4 | 100 | -1.018 | 0.290 | -1.127 | -0.910 | 0.084 |
|    | 200 | -0.977 | 0.159 | -1.037 | -0.918 | 0.026 |
|    | 500 | -0.989 | 0.128 | -1.036 | -0.941 | 0.016 |
|    | 1000 | -0.980 | 0.094 | -1.015 | -0.945 | 0.009 |
| R5 | 100 | -1.042 | 0.598 | -1.265 | -0.819 | 0.359 |
|    | 200 | -1.064 | 0.280 | -1.169 | -0.960 | 0.083 |
|    | 500 | -0.990 | 0.173 | -1.055 | -0.926 | 0.030 |
|    | 1000 | -0.969 | 0.143 | -1.023 | -0.916 | 0.021 |
| R6 | 100 | -1.077 | 0.507 | -1.267 | -0.888 | 0.263 |
|    | 200 | -0.993 | 0.351 | -1.125 | -0.862 | 0.124 |
|    | 500 | -0.952 | 0.212 | -1.031 | -0.873 | 0.047 |
|    | 1000 | -0.879 | 0.178 | -0.946 | -0.813 | 0.046 |

**Table 4:** Estimate of $\beta_1$ with true value of 1. Standard deviation of the random effects is 0.5.

| Run | N | Mean | SD | LCL | UCL | MSE |
|---|---|---|---|---|---|---|
| R1 | 100 | 1.072 | 0.175 | 1.007 | 1.137 | 0.036 |
|    | 200 | 0.997 | 0.138 | 0.945 | 1.048 | 0.019 |
|    | 500 | 1.007 | 0.085 | 0.975 | 1.039 | 0.007 |
|    | 1000 | 1.005 | 0.064 | 0.982 | 1.029 | 0.004 |
| R2 | 100 | 1.069 | 0.149 | 1.013 | 1.125 | 0.027 |
|    | 200 | 1.026 | 0.109 | 0.985 | 1.067 | 0.013 |
|    | 500 | 1.010 | 0.079 | 0.980 | 1.039 | 0.006 |
|    | 1000 | 0.992 | 0.053 | 0.973 | 1.012 | 0.003 |
| R3 | 100 | 1.069 | 0.167 | 1.007 | 1.131 | 0.033 |
|    | 200 | 1.013 | 0.177 | 0.947 | 1.079 | 0.032 |
|    | 500 | 1.030 | 0.101 | 0.993 | 1.068 | 0.011 |
|    | 1000 | 1.008 | 0.064 | 0.984 | 1.032 | 0.004 |
| R4 | 100 | 1.043 | 0.146 | 0.989 | 1.098 | 0.023 |
|    | 200 | 1.002 | 0.116 | 0.959 | 1.045 | 0.013 |
|    | 500 | 0.990 | 0.082 | 0.960 | 1.021 | 0.007 |
|    | 1000 | 0.978 | 0.057 | 0.957 | 0.999 | 0.004 |
| R5 | 100 | 1.007 | 0.180 | 0.940 | 1.074 | 0.032 |
|    | 200 | 1.035 | 0.128 | 0.987 | 1.083 | 0.018 |
|    | 500 | 1.000 | 0.119 | 0.955 | 1.044 | 0.014 |
|    | 1000 | 0.981 | 0.082 | 0.951 | 1.012 | 0.007 |
| R6 | 100 | 1.075 | 0.229 | 0.990 | 1.161 | 0.058 |
|    | 200 | 1.029 | 0.170 | 0.966 | 1.092 | 0.030 |
|    | 500 | 0.974 | 0.119 | 0.930 | 1.019 | 0.015 |
|    | 1000 | 0.953 | 0.089 | 0.920 | 0.986 | 0.010 |

**Table 5:** Estimate of $\beta_1$ with true value of 1. Standard deviation of the random effects is 1.

| Run | N | Mean | SD | LCL | UCL | MSE |
|---|---|---|---|---|---|---|
| R1 | 100 | 1.055 | 0.234 | 0.968 | 1.143 | 0.058 |
|  | 200 | 0.997 | 0.159 | 0.938 | 1.057 | 0.025 |
|  | 500 | 0.983 | 0.104 | 0.944 | 1.022 | 0.011 |
|  | 1000 | 0.991 | 0.073 | 0.963 | 1.018 | 0.005 |
| R2 | 100 | 1.063 | 0.192 | 0.992 | 1.135 | 0.041 |
|  | 200 | 1.018 | 0.144 | 0.965 | 1.072 | 0.021 |
|  | 500 | 0.973 | 0.092 | 0.939 | 1.007 | 0.009 |
|  | 1000 | 0.960 | 0.067 | 0.935 | 0.985 | 0.006 |
| R3 | 100 | 1.023 | 0.242 | 0.933 | 1.114 | 0.059 |
|  | 200 | 1.011 | 0.197 | 0.938 | 1.085 | 0.039 |
|  | 500 | 0.995 | 0.108 | 0.954 | 1.035 | 0.012 |
|  | 1000 | 0.974 | 0.084 | 0.943 | 1.005 | 0.008 |
| R4 | 100 | 1.062 | 0.208 | 0.984 | 1.139 | 0.047 |
|  | 200 | 1.011 | 0.154 | 0.954 | 1.069 | 0.024 |
|  | 500 | 0.993 | 0.098 | 0.956 | 1.029 | 0.010 |
|  | 1000 | 0.983 | 0.072 | 0.956 | 1.010 | 0.006 |
| R5 | 100 | 1.010 | 0.317 | 0.891 | 1.128 | 0.100 |
|  | 200 | 1.065 | 0.213 | 0.986 | 1.145 | 0.050 |
|  | 500 | 0.994 | 0.146 | 0.940 | 1.049 | 0.021 |
|  | 1000 | 0.987 | 0.097 | 0.951 | 1.023 | 0.010 |
| R6 | 100 | 1.131 | 0.303 | 1.017 | 1.244 | 0.109 |
|  | 200 | 1.075 | 0.224 | 0.991 | 1.159 | 0.056 |
|  | 500 | 0.980 | 0.145 | 0.926 | 1.034 | 0.022 |
|  | 1000 | 0.911 | 0.086 | 0.879 | 0.943 | 0.015 |

**Table 6:** Estimate of $\beta_1$ with true value of 1. Standard deviation of the random effects is 2.

| Run | N | Mean | SD | LCL | UCL | MSE |
|---|---|---|---|---|---|---|
| R1 | 100 | 1.065 | 0.415 | 0.910 | 1.220 | 0.176 |
|  | 200 | 0.998 | 0.269 | 0.898 | 1.099 | 0.072 |
|  | 500 | 0.971 | 0.177 | 0.904 | 1.037 | 0.032 |
|  | 1000 | 0.969 | 0.116 | 0.925 | 1.012 | 0.015 |
| R2 | 100 | 1.048 | 0.332 | 0.924 | 1.172 | 0.113 |
|  | 200 | 1.002 | 0.239 | 0.913 | 1.091 | 0.057 |
|  | 500 | 0.949 | 0.153 | 0.892 | 1.006 | 0.026 |
|  | 1000 | 0.914 | 0.114 | 0.872 | 0.957 | 0.020 |
| R3 | 100 | 0.973 | 0.442 | 0.807 | 1.138 | 0.197 |
|  | 200 | 0.996 | 0.291 | 0.887 | 1.105 | 0.085 |
|  | 500 | 0.956 | 0.180 | 0.889 | 1.024 | 0.034 |
|  | 1000 | 0.955 | 0.138 | 0.903 | 1.006 | 0.021 |
| R4 | 100 | 1.113 | 0.368 | 0.975 | 1.250 | 0.148 |
|  | 200 | 1.044 | 0.273 | 0.942 | 1.146 | 0.077 |
|  | 500 | 0.988 | 0.155 | 0.930 | 1.046 | 0.024 |
|  | 1000 | 0.979 | 0.126 | 0.932 | 1.026 | 0.016 |
| R5 | 100 | 1.129 | 0.752 | 0.849 | 1.410 | 0.582 |
|  | 200 | 1.141 | 0.364 | 1.005 | 1.277 | 0.152 |
|  | 500 | 0.970 | 0.261 | 0.872 | 1.067 | 0.069 |
|  | 1000 | 0.980 | 0.159 | 0.920 | 1.039 | 0.026 |
| R6 | 100 | 1.137 | 0.474 | 0.960 | 1.314 | 0.243 |
|  | 200 | 1.076 | 0.309 | 0.960 | 1.191 | 0.101 |
|  | 500 | 0.970 | 0.187 | 0.900 | 1.040 | 0.036 |
|  | 1000 | 0.868 | 0.155 | 0.810 | 0.926 | 0.041 |

**Table 7:** Estimate of $\beta_2$ with true value of 1. Standard deviation of the random effects is 0.5.

| Run | N | Mean | SD | LCL | UCL | MSE |
|---|---|---|---|---|---|---|
| R1 | 100 | 1.039 | 0.108 | 0.999 | 1.079 | 0.013 |
|  | 200 | 1.011 | 0.102 | 0.973 | 1.049 | 0.011 |
|  | 500 | 1.031 | 0.057 | 1.009 | 1.052 | 0.004 |
|  | 1000 | 1.028 | 0.038 | 1.014 | 1.042 | 0.002 |
| R2 | 100 | 1.031 | 0.091 | 0.997 | 1.065 | 0.009 |
|  | 200 | 1.008 | 0.076 | 0.980 | 1.036 | 0.006 |
|  | 500 | 1.020 | 0.043 | 1.004 | 1.036 | 0.002 |
|  | 1000 | 1.015 | 0.030 | 1.004 | 1.027 | 0.001 |
| R3 | 100 | 1.037 | 0.123 | 0.992 | 1.083 | 0.016 |
|  | 200 | 0.994 | 0.137 | 0.943 | 1.045 | 0.019 |
|  | 500 | 1.028 | 0.069 | 1.003 | 1.054 | 0.006 |
|  | 1000 | 1.023 | 0.046 | 1.006 | 1.041 | 0.003 |
| R4 | 100 | 0.984 | 0.091 | 0.951 | 1.018 | 0.008 |
|  | 200 | 0.970 | 0.079 | 0.940 | 0.999 | 0.007 |
|  | 500 | 0.985 | 0.044 | 0.969 | 1.001 | 0.002 |
|  | 1000 | 0.982 | 0.029 | 0.971 | 0.993 | 0.001 |
| R5 | 100 | 0.978 | 0.148 | 0.922 | 1.033 | 0.022 |
|  | 200 | 0.983 | 0.110 | 0.942 | 1.025 | 0.012 |
|  | 500 | 0.997 | 0.065 | 0.972 | 1.021 | 0.004 |
|  | 1000 | 0.980 | 0.041 | 0.965 | 0.996 | 0.002 |
| R6 | 100 | 0.955 | 0.178 | 0.888 | 1.021 | 0.034 |
|  | 200 | 0.959 | 0.159 | 0.900 | 1.019 | 0.027 |
|  | 500 | 0.960 | 0.096 | 0.925 | 0.996 | 0.011 |
|  | 1000 | 0.963 | 0.063 | 0.939 | 0.986 | 0.005 |

**Table 8:** Estimate of $\beta_2$ with true value of 1. Standard deviation of the random effects is 1.

| Run | N | Mean | SD | LCL | UCL | MSE |
|---|---|---|---|---|---|---|
| R1 | 100 | 0.976 | 0.112 | 0.934 | 1.018 | 0.013 |
|  | 200 | 0.974 | 0.095 | 0.938 | 1.009 | 0.010 |
|  | 500 | 0.992 | 0.072 | 0.965 | 1.019 | 0.005 |
|  | 1000 | 0.986 | 0.033 | 0.973 | 0.998 | 0.001 |
| R2 | 100 | 0.972 | 0.099 | 0.935 | 1.009 | 0.011 |
|  | 200 | 0.960 | 0.067 | 0.935 | 0.985 | 0.006 |
|  | 500 | 0.969 | 0.050 | 0.950 | 0.987 | 0.003 |
|  | 1000 | 0.969 | 0.031 | 0.958 | 0.981 | 0.002 |
| R3 | 100 | 0.964 | 0.171 | 0.900 | 1.028 | 0.031 |
|  | 200 | 0.941 | 0.137 | 0.890 | 0.992 | 0.022 |
|  | 500 | 0.982 | 0.081 | 0.852 | 1.012 | 0.007 |
|  | 1000 | 0.975 | 0.049 | 0.957 | 0.993 | 0.003 |
| R4 | 100 | 0.982 | 0.105 | 0.943 | 1.025 | 0.011 |
|  | 200 | 0.970 | 0.069 | 0.944 | 0.996 | 0.006 |
|  | 500 | 0.986 | 0.046 | 0.969 | 1.003 | 0.002 |
|  | 1000 | 0.981 | 0.033 | 0.969 | 0.993 | 0.001 |
| R5 | 100 | 0.956 | 0.143 | 0.903 | 1.010 | 0.022 |
|  | 200 | 0.988 | 0.112 | 0.946 | 1.029 | 0.013 |
|  | 500 | 0.996 | 0.085 | 0.965 | 1.028 | 0.007 |
|  | 1000 | 0.980 | 0.042 | 0.965 | 0.996 | 0.002 |
| R6 | 100 | 0.940 | 0.214 | 0.860 | 1.020 | 0.049 |
|  | 200 | 0.955 | 0.155 | 0.897 | 1.013 | 0.026 |
|  | 500 | 0.953 | 0.097 | 0.917 | 0.990 | 0.012 |
|  | 1000 | 0.920 | 0.082 | 0.889 | 0.950 | 0.013 |

**Table 9:** Estimate of $\beta_2$ with true value of 1. Standard deviation of the random effects is 2.

| Run | N | Mean | SD | LCL | UCL | MSE |
|---|---|---|---|---|---|---|
| R1 | 100 | 0.933 | 0.158 | 0.874 | 0.992 | 0.029 |
|  | 200 | 0.956 | 0.115 | 0.913 | 0.999 | 0.015 |
|  | 500 | 0.983 | 0.089 | 0.950 | 1.016 | 0.008 |
|  | 1000 | 0.967 | 0.044 | 0.951 | 0.984 | 0.003 |
| R2 | 100 | 0.955 | 0.160 | 0.895 | 1.014 | 0.028 |
|  | 200 | 0.933 | 0.079 | 0.903 | 0.962 | 0.011 |
|  | 500 | 0.941 | 0.066 | 0.916 | 0.966 | 0.008 |
|  | 1000 | 0.933 | 0.043 | 0.917 | 0.949 | 0.006 |
| R3 | 100 | 0.978 | 0.265 | 0.880 | 1.077 | 0.071 |
|  | 200 | 0.914 | 0.189 | 0.843 | 0.984 | 0.043 |
|  | 500 | 0.957 | 0.114 | 0.915 | 1.000 | 0.015 |
|  | 1000 | 0.944 | 0.068 | 0.919 | 0.970 | 0.008 |
| R4 | 100 | 0.977 | 0.135 | 0.927 | 1.027 | 0.019 |
|  | 200 | 0.962 | 0.067 | 0.938 | 0.987 | 0.006 |
|  | 500 | 0.986 | 0.068 | 0.961 | 1.012 | 0.005 |
|  | 1000 | 0.992 | 0.039 | 0.977 | 1.006 | 0.002 |
| R5 | 100 | 0.943 | 0.194 | 0.870 | 1.015 | 0.041 |
|  | 200 | 0.988 | 0.137 | 0.936 | 1.039 | 0.019 |
|  | 500 | 1.003 | 0.095 | 0.967 | 1.038 | 0.009 |
|  | 1000 | 0.947 | 0.129 | 0.899 | 0.996 | 0.019 |
| R6 | 100 | 1.014 | 0.271 | 0.912 | 1.115 | 0.074 |
|  | 200 | 0.976 | 0.186 | 0.906 | 1.046 | 0.035 |
|  | 500 | 0.947 | 0.129 | 0.899 | 0.996 | 0.019 |
|  | 1000 | 0.904 | 0.112 | 0.862 | 0.946 | 0.022 |

**Table 10:** Estimate of $\sigma$ with true value of 0.5.

| Run | N | Mean | SD | LCL | UCL | MSE |
|---|---|---|---|---|---|---|
| R1 | 100 | 0.869 | 0.054 | 0.849 | 0.889 | 0.139 |
|  | 200 | 0.884 | 0.036 | 0.871 | 0.898 | 0.149 |
|  | 500 | 0.877 | 0.028 | 0.866 | 0.887 | 0.143 |
|  | 1000 | 0.874 | 0.020 | 0.866 | 0.881 | 0.140 |
| R2 | 100 | 0.917 | 0.054 | 0.897 | 0.937 | 0.177 |
|  | 200 | 0.919 | 0.033 | 0.907 | 0.931 | 0.177 |
|  | 500 | 0.923 | 0.025 | 0.913 | 0.932 | 0.179 |
|  | 1000 | 0.932 | 0.019 | 0.925 | 0.939 | 0.187 |
| R3 | 100 | 0.923 | 0.082 | 0.893 | 0.954 | 0.186 |
|  | 200 | 0.933 | 0.057 | 0.911 | 0.954 | 0.190 |
|  | 500 | 0.917 | 0.033 | 0.905 | 0.929 | 0.175 |
|  | 1000 | 0.921 | 0.027 | 0.911 | 0.931 | 0.178 |
| R4 | 100 | 0.435 | 0.132 | 0.386 | 0.485 | 0.022 |
|  | 200 | 0.463 | 0.084 | 0.432 | 0.495 | 0.008 |
|  | 500 | 0.475 | 0.061 | 0.452 | 0.498 | 0.004 |
|  | 1000 | 0.433 | 0.075 | 0.405 | 0.461 | 0.010 |
| R5 | 100 | 0.440 | 0.175 | 0.375 | 0.505 | 0.034 |
|  | 200 | 0.461 | 0.133 | 0.411 | 0.510 | 0.019 |
|  | 500 | 0.491 | 0.073 | 0.464 | 0.519 | 0.005 |
|  | 1000 | 0.447 | 0.082 | 0.416 | 0.477 | 0.010 |
| R6 | 100 | 0.385 | 0.179 | 0.318 | 0.451 | 0.045 |
|  | 200 | 0.369 | 0.173 | 0.304 | 0.433 | 0.047 |
|  | 500 | 0.298 | 0.196 | 0.225 | 0.371 | 0.079 |
|  | 1000 | 0.210 | 0.141 | 0.157 | 0.262 | 0.104 |

**Table 11:** Estimate of σ with true value of 1.

| Run | N | Mean | SD | LCL | UCL | MSE |
|---|---|---|---|---|---|---|
| R1 | 100 | 1.015 | 0.149 | 0.960 | 1.071 | 0.022 |
|    | 200 | 1.007 | 0.084 | 0.976 | 1.039 | 0.007 |
|    | 500 | 1.009 | 0.055 | 0.988 | 1.029 | 0.003 |
|    | 1000 | 0.997 | 0.040 | 0.982 | 1.012 | 0.002 |
| R2 | 100 | 0.999 | 0.101 | 0.962 | 1.037 | 0.010 |
|    | 200 | 1.004 | 0.066 | 0.979 | 1.029 | 0.004 |
|    | 500 | 1.018 | 0.041 | 1.003 | 1.034 | 0.002 |
|    | 1000 | 1.022 | 0.026 | 1.012 | 1.032 | 0.001 |
| R3 | 100 | 1.038 | 0.190 | 0.967 | 1.109 | 0.038 |
|    | 200 | 1.038 | 0.134 | 0.987 | 1.088 | 0.019 |
|    | 500 | 1.006 | 0.057 | 0.984 | 1.027 | 0.003 |
|    | 1000 | 1.012 | 0.049 | 0.994 | 1.030 | 0.003 |
| R4 | 100 | 0.992 | 0.139 | 0.940 | 1.044 | 0.019 |
|    | 200 | 0.984 | 0.095 | 0.948 | 1.019 | 0.009 |
|    | 500 | 1.004 | 0.060 | 0.982 | 1.027 | 0.004 |
|    | 1000 | 0.972 | 0.069 | 0.946 | 0.998 | 0.006 |
| R5 | 100 | 1.061 | 0.205 | 0.984 | 1.138 | 0.046 |
|    | 200 | 1.004 | 0.147 | 0.949 | 1.059 | 0.022 |
|    | 500 | 1.011 | 0.088 | 0.978 | 1.044 | 0.008 |
|    | 1000 | 0.984 | 0.063 | 0.961 | 1.007 | 0.004 |
| R6 | 100 | 0.949 | 0.223 | 0.866 | 1.032 | 0.052 |
|    | 200 | 0.962 | 0.158 | 0.904 | 1.021 | 0.026 |
|    | 500 | 0.894 | 0.168 | 0.831 | 0.957 | 0.040 |
|    | 1000 | 0.644 | 0.266 | 0.544 | 0.743 | 0.198 |

**Table 12:** Estimate of σ with true value of 2.

| Run | N | Mean | SD | LCL | UCL | MSE |
|---|---|---|---|---|---|---|
| R1 | 100 | 1.857 | 0.442 | 1.692 | 2.022 | 0.216 |
|    | 200 | 1.880 | 0.254 | 1.785 | 1.975 | 0.079 |
|    | 500 | 1.849 | 0.163 | 1.788 | 1.910 | 0.050 |
|    | 1000 | 1.848 | 0.113 | 1.806 | 1.890 | 0.036 |
| R2 | 100 | 1.787 | 0.275 | 1.685 | 1.890 | 0.121 |
|    | 200 | 1.796 | 0.212 | 1.717 | 1.875 | 0.087 |
|    | 500 | 1.749 | 0.151 | 1.693 | 1.806 | 0.086 |
|    | 1000 | 1.783 | 0.139 | 1.731 | 1.834 | 0.067 |
| R3 | 100 | 1.806 | 0.472 | 1.630 | 1.982 | 0.261 |
|    | 200 | 1.878 | 0.294 | 1.768 | 1.988 | 0.101 |
|    | 500 | 1.761 | 0.202 | 1.686 | 1.837 | 0.098 |
|    | 1000 | 1.743 | 0.160 | 1.683 | 1.802 | 0.092 |
| R4 | 100 | 2.052 | 0.220 | 1.970 | 2.135 | 0.051 |
|    | 200 | 2.009 | 0.165 | 1.947 | 2.070 | 0.027 |
|    | 500 | 2.006 | 0.111 | 1.964 | 2.047 | 0.012 |
|    | 1000 | 2.009 | 0.076 | 1.981 | 2.037 | 0.006 |
| R5 | 100 | 2.178 | 0.389 | 2.033 | 2.323 | 0.183 |
|    | 200 | 2.066 | 0.265 | 1.967 | 2.165 | 0.074 |
|    | 500 | 2.016 | 0.150 | 1.960 | 2.072 | 0.023 |
|    | 1000 | 1.998 | 0.107 | 1.958 | 2.038 | 0.011 |
| R6 | 100 | 2.082 | 0.279 | 1.977 | 2.186 | 0.085 |
|    | 200 | 2.024 | 0.217 | 1.943 | 2.105 | 0.048 |
|    | 500 | 1.895 | 0.210 | 1.816 | 1.973 | 0.055 |
|    | 1000 | 1.717 | 0.293 | 1.607 | 1.826 | 0.166 |

**Table 13:** Estimate of parameters with *T=100, I=100, N=30,* and $\sigma = 1$.

| Run | Parameter | Mean | SD | LCL | UCL | MSE |
|---|---|---|---|---|---|---|
| R2 | $\beta_0$ | -1.055 | 0.181 | -1.123 | -0.987 | 0.036 |
|  | $\beta_1$ | 1.018 | 0.241 | 0.928 | 1.108 | 0.058 |
|  | $\beta_2$ | 1.08 | 0.168 | 1.018 | 1.143 | 0.035 |
|  | $\sigma$ | 1.036 | 0.121 | 0.991 | 1.081 | 0.016 |
| R6 | $\beta_0$ | 9.002 | 11.931 | 4.547 | 13.457 | 242.389 |
|  | $\beta_1$ | 2.026 | 21.951 | -6.17 | 10.223 | 482.899 |
|  | $\beta_2$ | 74.863 | 24.809 | 65.599 | 84.127 | 6071.229 |
|  | $\sigma$ | 1.163 | 1.892 | 0.457 | 1.87 | 3.606 |

**Table 14:** Estimate of parameters for SP index data.

| Run | Parameter | Mean | SD | LCL | UCL |
|---|---|---|---|---|---|
| Uninformative Prior | $\beta_0$ | -0.180 | 4.081 | -9.457 | 6.927 |
|  | $\beta_1$ | -0.582 | 0.804 | -2.074 | 1.154 |
|  | $\sigma$ | 1.061 | 0.284 | 0.566 | 1.619 |
| Informative prior | $\beta_0$ | -4.447 | 0.288 | -5.024 | -3.865 |
|  | $\beta_1$ | 0.297 | 0.072 | 0.158 | 0.437 |
|  | $\sigma$ | 1.053 | 0.256 | 0.640 | 1.590 |

**Acknowledgement:** The author appreciates the effort of Norita Dobyns in writing the SAS programs during her work-study at the University of the Fraser Valley.